\documentclass[prb,twocolumn,english,superscriptaddress,floatfix]{revtex4-1}

\usepackage{graphicx}
\usepackage{dcolumn}
\usepackage{color}
\usepackage{array,multirow,makecell}
\usepackage{bm}
\usepackage{nicefrac}

\begin{document}
\title{Magnetic reshuffling and feedback on superconductivity in UTe$_2$ under pressure}

\author{M. Vali\v{s}ka}
\affiliation{Univ. Grenoble Alpes, CEA, Grenoble INP, IRIG, PHELIQS, 38000, Grenoble, France}
\affiliation{Charles University, Faculty of Mathematics and Physics, Department of Condensed Matter Physics, Ke Karlovu 5, Prague 2, 121 16, Czech Republic }

\author{W. Knafo}
\affiliation{Laboratoire National des Champs Magn\'etiques Intenses - EMFL, CNRS, Univ. Grenoble Alpes, INSA-T, Univ. Toulouse 3, 31400 Toulouse, France}

\author{G. Knebel}
\affiliation{Univ. Grenoble Alpes, CEA, Grenoble INP, IRIG, PHELIQS, 38000, Grenoble, France}

\author{G. Lapertot}
\affiliation{Univ. Grenoble Alpes, CEA, Grenoble INP, IRIG, PHELIQS, 38000, Grenoble, France}

\author{D. Aoki}
\affiliation{Institute for Materials Research, Tohoku University, Ibaraki 311-1313, Japan}

\author{D. Braithwaite}
\affiliation{Univ. Grenoble Alpes, CEA, Grenoble INP, IRIG, PHELIQS, 38000, Grenoble, France}

%

\date{\today}

\begin{abstract}
The discovery of superconductivity in the heavy-fermion paramagnet UTe$_2$ has attracted a lot of attention, particularly due to the reinforcement of superconductivity near pressure- and magnetic-field-induced magnetic quantum phase transitions. A challenge is now to characterize the effects of combined pressure and magnetic fields applied along variable directions in this strongly anisotropic paramagnet. Here, we present an investigation of the electrical resistivity of UTe$_2$ under pressure up to 3~GPa and pulsed magnetic fields up to 58~T along the hard magnetic crystallographic directions $\mathbf{b}$ and $\mathbf{c}$. We construct three-dimensional phase diagrams and show that, near the critical pressure, a field-enhancement of superconductivity coincides with a boost of the effective mass related to the collapse of metamagnetic and critical fields at the boundaries of the correlated paramagnetic regime and magnetically-ordered phase, respectively. Beyond the critical pressure, field-induced transitions precede the destruction of the magnetically-ordered phase, suggesting an antiferromagnetic nature. By bringing new elements about the interplay between magnetism and superconductivity, our work appeals for microscopic theories describing the anisotropic properties of UTe$_2$ under pressure and magnetic field.

\end{abstract}

\maketitle

\section{Introduction}

The recent discovery of superconductivity in the strongly correlated system UTe$_2$ has sparked enormous interest \cite{Ran2019,Aoki2019b}. This orthorhombic compound is a paramagnet with anisotropic magnetic properties \cite{Ikeda2006,Ran2019,Aoki2019b,Miyake2019,Miyake2021}: the magnetic susceptibility along the $a$-axis increases strongly at low temperature, leading to the initial suggestion that the system is very close to ferromagnetic order \cite{Ran2019}, whereas the other directions are "hard" magnetization axes, with $\mathbf{b}$ being the hardest at low temperature. But the properties of the superconducting state are the most striking aspect, and in particular the strong enhancement of superconductivity when a magnetic field $\mathbf{H}$ is applied along the $b$-axis \cite{Ran2019b,Knebel2019}. In this case superconductivity persists in magnetic fields up to $\mu_0H_m=35$~T, where a first order metamagnetic transition occurs with a large jump of the magnetization \cite{Miyake2019}, a similarly large jump in the residual electrical resistivity of the normal state \cite{Knafo2019}, and the destruction of superconductivity \cite{Knebel2019,Ran2019b,Niu2020b}. Even more remarkably when the field is tilted by about $30~^\circ$ from the $b$-axis in the hard $\mathbf{b}-\mathbf{c}$ plane, superconductivity re-emerges above $\mu_0H_m \simeq 40$~T for this angle \cite{Ran2019b,Knafo2021}. The extremely high values of the upper critical field, $H_{c2}$, compared to the initial superconducting critical temperature ($T_{sc} = 1.6$~K) suggest a probable spin-triplet order parameter, at least in some parts of the phase diagram. This enhancement of superconductivity is very reminiscent of the phenomenon found in the ferromagnetic superconductors URhGe \cite{Levy2005} and UCoGe \cite{Aoki2009}. However, in these cases the reinforcement of superconductivity, when a field is applied along a hard magnetic axis, is understood as a consequence of the collapse of ferromagnetism, since an enhancement of the ferromagnetic fluctuations have been shown to be responsible for the superconducting pairing \cite{Wu2017,Tokunaga2015}. This explanation can obviously not be directly transposed to UTe$_2$ where no sign of magnetic ordering has been found down to very low temperatures \cite{Sundar2019,Paulsen2021}. Low-dimensional antiferromagnetic fluctuations were reported, suggesting that UTe$_2$, whose U ions form a magnetic ladder structure, is subject to antiferromagnetic exchange leading to antiferromagnetic correlations \cite{Duan2020,Knafo2021b}. The opening of a gap associated with these antiferromagnetic fluctuations was also observed in the superconducting phase  \cite{Duan2021,Raymond2021}. The magnetic properties of UTe$_2$ are thus associated to its unusual superconducting properties. A full description of the relationship between the two is essential to understand superconductivity in UTe$_2$, and may well advance our understanding of magnetically-mediated superconductivity in general.

Applying pressure is the tool of choice to tune magnetism in strongly correlated systems. Often pressure ($p$) can drive a system towards and through a magnetic instability, giving a direct probe of the relationship between magnetism and superconductivity. For UTe$_2$ it has already been shown that hydrostatic pressure induces an enhancement of $T_{sc}$ by a factor 2, reaching about 3~K \cite{Braithwaite2019,Knebel2020,Ran2020,Thomas2020}. Pressure has also revealed further complexities of this system's superconducting state, with multiple superconducting order parameters appearing \cite{Braithwaite2019,Thomas2020,Aoki2020,Lin2020}. It was also shown that above a critical pressure $p_c\simeq1.5-1.7$~GPa, magnetic order occurs with the concomitant disappearance of superconductivity \cite{Braithwaite2019,Knebel2020,Ran2020,Thomas2020,Aoki2020,Lin2020}. Furthermore the metamagnetic field decreases strongly with pressure for $\mathbf{H}\parallel\mathbf{b}$ \cite{Knebel2020}, and when field is applied along the $a$-axis (the easy axis at ambient pressure) the multiple superconducting states have quite different behaviours \cite{Aoki2020}. A large enhancement of $H_{c2}$ for $\mathbf{H}\parallel\mathbf{c}$ was also found with an inversion of the anisotropy of $H_{c2}$ in the $\mathbf{b}-\mathbf{c}$ plane for pressures close to $p_c$ \cite{Knebel2020}. A recent study showed that the magnetic anisotropy is also significantly changed with pressure, with $\mathbf{b}$ becoming the easy axis above $p_c$ \cite{Li2021}. In addition, re-entrant superconductivity occurs for $\mathbf{H}\parallel\mathbf{c}$ just above the critical pressure \cite{Aoki2021}. The above studies were mainly performed in static magnetic fields, but the highest static fields available, even in dedicated facilities, are insufficient to reveal all the physics in UTe$_2$. Here we report on an experiment combining high pressure and pulsed magnetic fields up to 60~T. These measurements were also performed over a wide temperature range giving new insight into the evolution of the magnetic properties and their feedback on superconductivity in three-dimensional (3D) $(H,p,T)$ phase diagrams of UTe$_2$.

\begin{figure}[t]
\includegraphics[width=1\columnwidth]{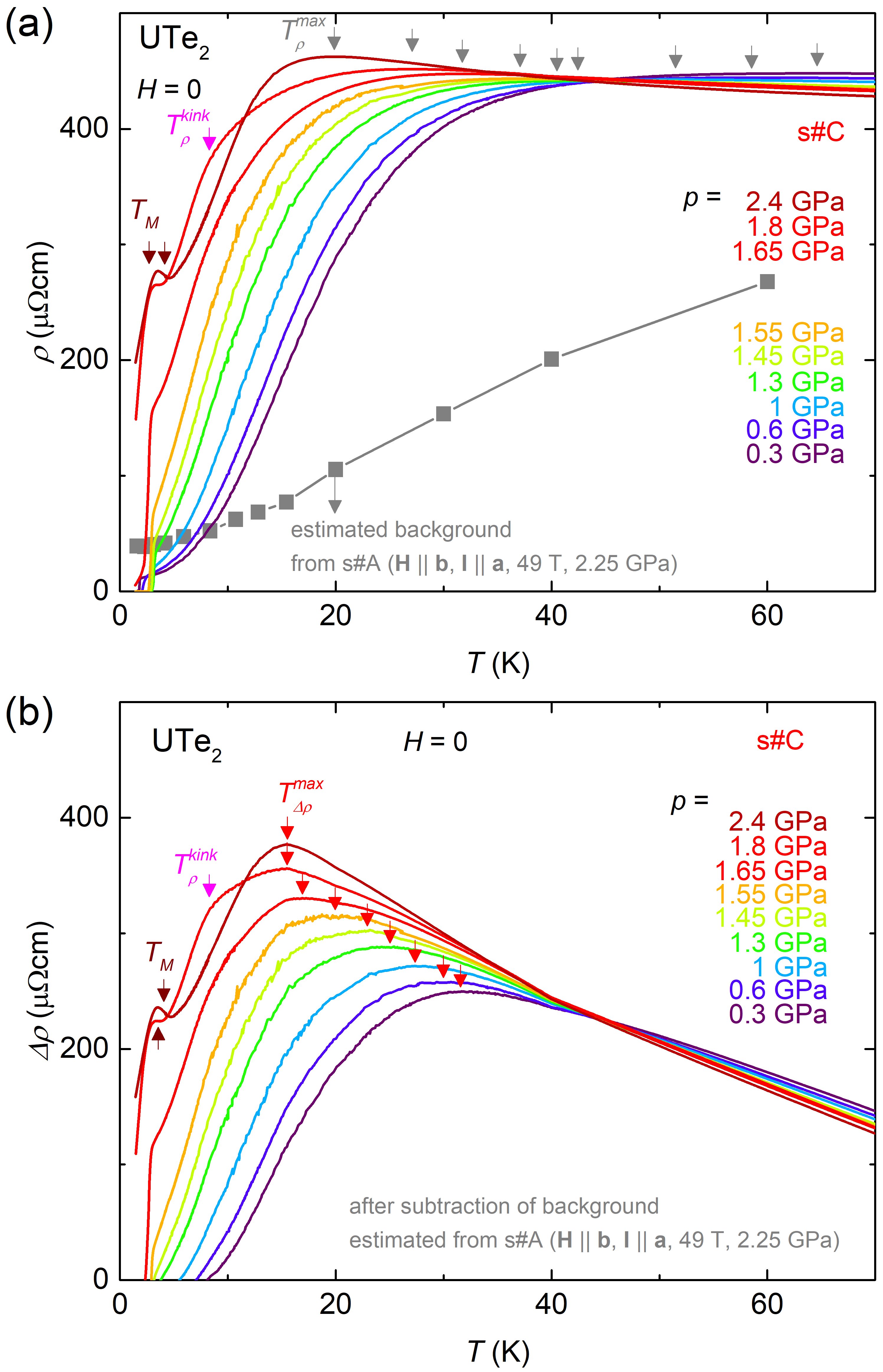}
\caption{(a) Zero-field temperature-dependence of the electrical resistivity $\rho$ of UTe$_2$ under pressure. The grey squares show the curve at high pressure and high field that we take as a background. (b) Temperature dependence of resistivity $\Delta\rho$ determined after subtraction of the background.}
\label{Fig1}
\end{figure}

\section{Methods}

We used a previously described pressure cell \cite{Braithwaite2016} allowing magnetoresistivity measurements in pulsed magnetic fields up to 60~T and temperatures down to 1.4~K. Single crystals of UTe$_2$ were grown by the chemical vapor transport technique as described elsewhere \cite{Aoki2019b}. The crystals were oriented by X-ray Laue diffraction and cut to bar shaped samples of about $0.8\times0.2\times0.1$~mm$^3$. Two successive experiments were performed in a pressure cell, offering the simultaneous measurement of the electrical resistivity of two samples in a magnetic field along the $\mathbf{b}$ and $\mathbf{c}$ directions. First experiment was performed at pressures above $p_c$ on samples $\sharp$A with $\mathbf{H}\parallel\mathbf{b}$ and $\sharp$B with $\mathbf{H}\parallel\mathbf{c}$ (see Supplementary Materials \cite{SM}). Second experiment was performed under a large set of pressures below and above $p_c$, from 0.3 to 3.1~GPa, on samples $\sharp$C with $\mathbf{H}\parallel\mathbf{b}$ and $\sharp$D with $\mathbf{H}\parallel\mathbf{c}$, and corresponds to the data presented in this manuscript (see also Supplementary Materials \cite{SM}). Both experiments gave similar results, although the samples set up for $\mathbf{H}\parallel\mathbf{b}$ displayed an unintentional misalignment (see Section \ref{results}), probably having moved on pressurization. A piece of lead was also mounted in the cell to determine the pressure. High-pressure magnetoresistivity measurements were performed at the Laboratoire National des Champs Magn\'{e}tiques Intenses (LNCMI) in Toulouse under long-duration (50~ms rise and 300~ms fall) pulsed magnetic fields up to 58~T and temperatures down to 1.4~K. A standard four-probe method with a current $\mathbf{I}\parallel\mathbf{a}$ of 0.5~mA, at a frequency of 15-70~kHz and digital lock-in detection was used. The temperature dependence of the resistivity was also measured directly in zero field.

\section{Results}
\label{results}

\subsection{Zero-field high-pressure properties}

\begin{figure}[t]
\includegraphics[width=1\columnwidth]{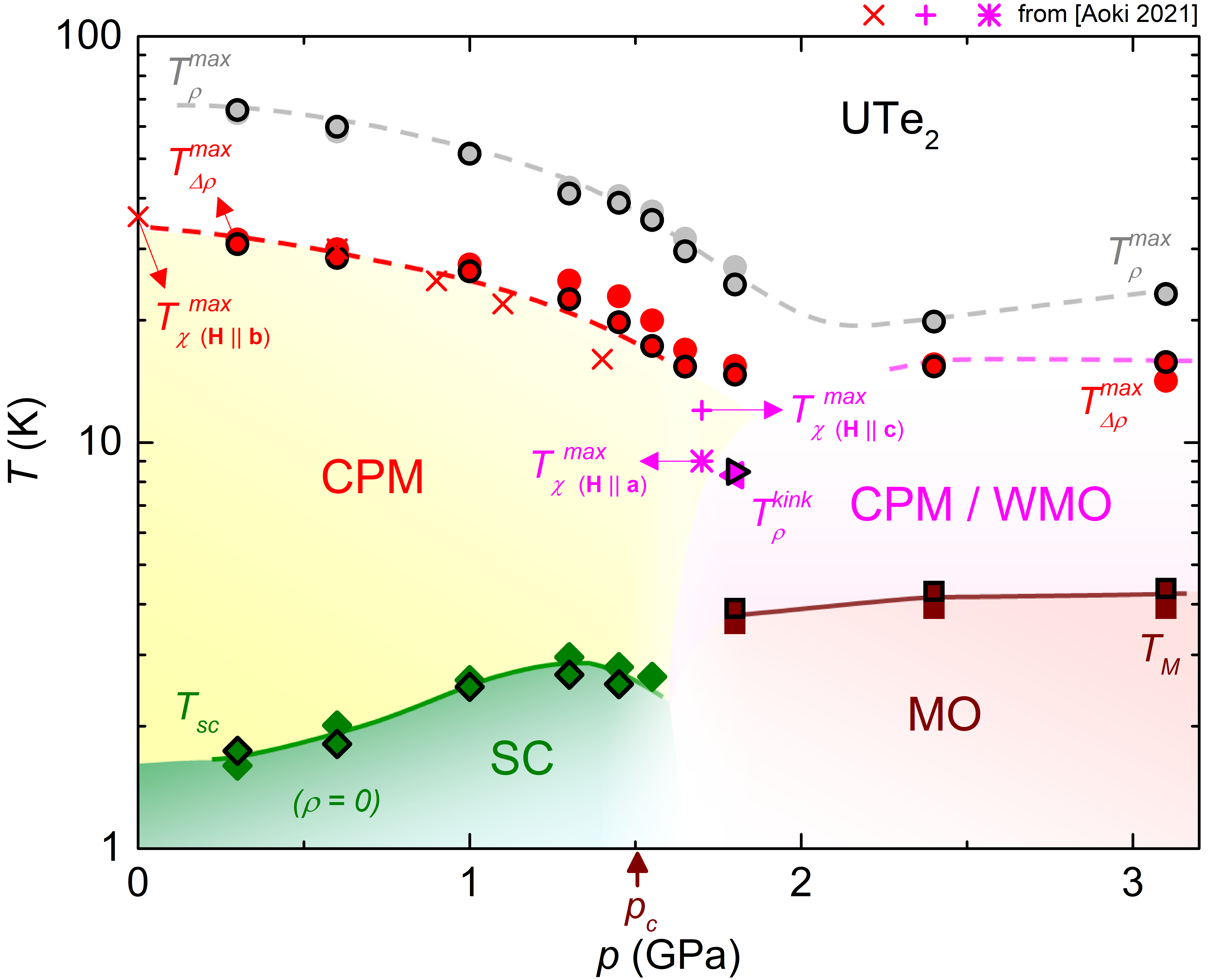}
\caption{Phase diagram of UTe$_2$ from these and previous \cite{Aoki2021} measurements. CPM and CPM/WMO denote the correlated paramagnetic regimes stabilized at pressures below and above $p_c$, respectively, SC the superconducting phase, and MO the magnetically-ordered phase.}
\label{Fig2}
\end{figure}

Zero-field electrical resistivity $\rho(T)$ curves measured at different pressures are shown in Fig. \ref{Fig1}(a-b). Below $p_c$, the onset of superconductivity is visible at low temperatures, leading to zero resistivity. Above $p_c$ a different anomaly appears in the resistivity curve (labeled $T_M$) that is almost certainly the signature of long range magnetic order. In the 1.8~GPa curve a further anomaly is apparent at higher temperature ($T_\rho^{kink}$). A similar anomaly has been seen in other resistivity studies at pressures just above $p_c$ \cite{Thomas2020} and a broad anomaly is also visible in the magnetization \cite{Li2021}. Another feature of the $\rho(T)$ curves is a broad maximum, which occurs at a temperature $T_\rho^{max}\simeq60$~K at ambient pressure. $T_\rho^{max}$ decreases with pressure down to about 20~K at $p_c$ and then slightly shifts to higher temperatures again as pressure is increased above $p_c$. This maximum is a general feature in heavy fermion systems and is an indication of the energy scale below which the cross-over between a high temperature paramagnetic state (PM) to a coherent heavy fermion state or correlated paramagnetic state (CPM) occurs.

In Fig. \ref{Fig2} we show the zero-field pressure-temperature phase diagram constructed from our measurements together with data from previous studies \cite{Aoki2021}. The superconducting critical temperature $T_{sc}$ increases from an initial value of 1.7~K up to a maximum value of about 3~K at a pressure of about 1.3~GPa. $T_{sc}$ then decreases abruptly and disappears at a critical pressure $p_c\simeq1.5$~GPa. Reported values of $p_c$ vary, probably due to different pressure conditions, ranging from about 1.4 to 1.7~GPa \cite{Braithwaite2019,Knebel2020,Ran2020,Thomas2020,Aoki2020,Lin2020,Aoki2021}. An interesting feature is the maximum in the susceptibility $T_{\chi(\mathbf{H}\parallel\mathbf{b})}^{max}$ observed for $\mathbf{H}\parallel\mathbf{b}$ below $p_c$ and for $\mathbf{H}\parallel\mathbf{a},\mathbf{c}$ above $p_c$ \cite{Ikeda2006,Ran2019,Knebel2020,Aoki2021}. $T_\rho^{max}$ and $T_{\chi(\mathbf{H}\parallel\mathbf{b})}^{max}$ show similar behavior with pressure up to $p_c$, although their values are quite different. Of course, neither feature is a precise indication of this energy scale which is anyway a cross-over, but the large temperature difference here is due to the effect of other contributions to the resistivity, including (but not limited to) phonon scattering, which should be subtracted to get the magnetic scattering. The temperature shift induced by this effect can be quite significant in the case of UTe$_2$, as the maximum is rather broad and weak. To check, we subtracted the background shown in Fig. \ref{Fig1}(a), corresponding to a $\rho(T)$ curve obtained under combined high pressure and magnetic field. This background corresponds to a high-field regime where the ground state is magnetically polarized, and for which a significant part of the magnetic correlations has been suppressed. Empirically we see that, once the background is subtracted, the maximum is much more pronounced and occurs at a temperature $T_{\Delta\rho}^{max}$ quite close to $T_{\chi(\mathbf{H}\parallel\mathbf{b})}^{max}$, which delimitates a CPM regime [see  Fig. \ref{Fig1}(b)]. Beyond the critical pressure, a switch of the magnetic properties is observed, in relation with the onset of long-range magnetic ordering (MO), and possibly higher-temperature correlated paramagnetism or short-range weak magnetic ordering (noted CPM/WMO, see later): a maximum in the magnetic susceptibility is observed for $\mathbf{H}\parallel\mathbf{a},\mathbf{c}$ at the temperatures $T_{\chi(\mathbf{H}\parallel\mathbf{a})}^{max}$ and $T_{\chi(\mathbf{H}\parallel\mathbf{c})}^{max}$, respectively, but no maximum of the magnetic susceptibility is observed for $\mathbf{H}\parallel\mathbf{b}$ \cite{Li2021}. Knowing that it is also possible to follow the maximum in $\Delta\rho$ to high fields as will be shown later, we therefore conclude that $T_{\Delta\rho}^{max}$ is a good criterion to follow the cross-over to the CPM regime as a function of pressure, field and temperature.

\subsection{High-field and high-pressure electrical resistivity}

\begin{figure}[h]
\includegraphics[width=1\columnwidth]{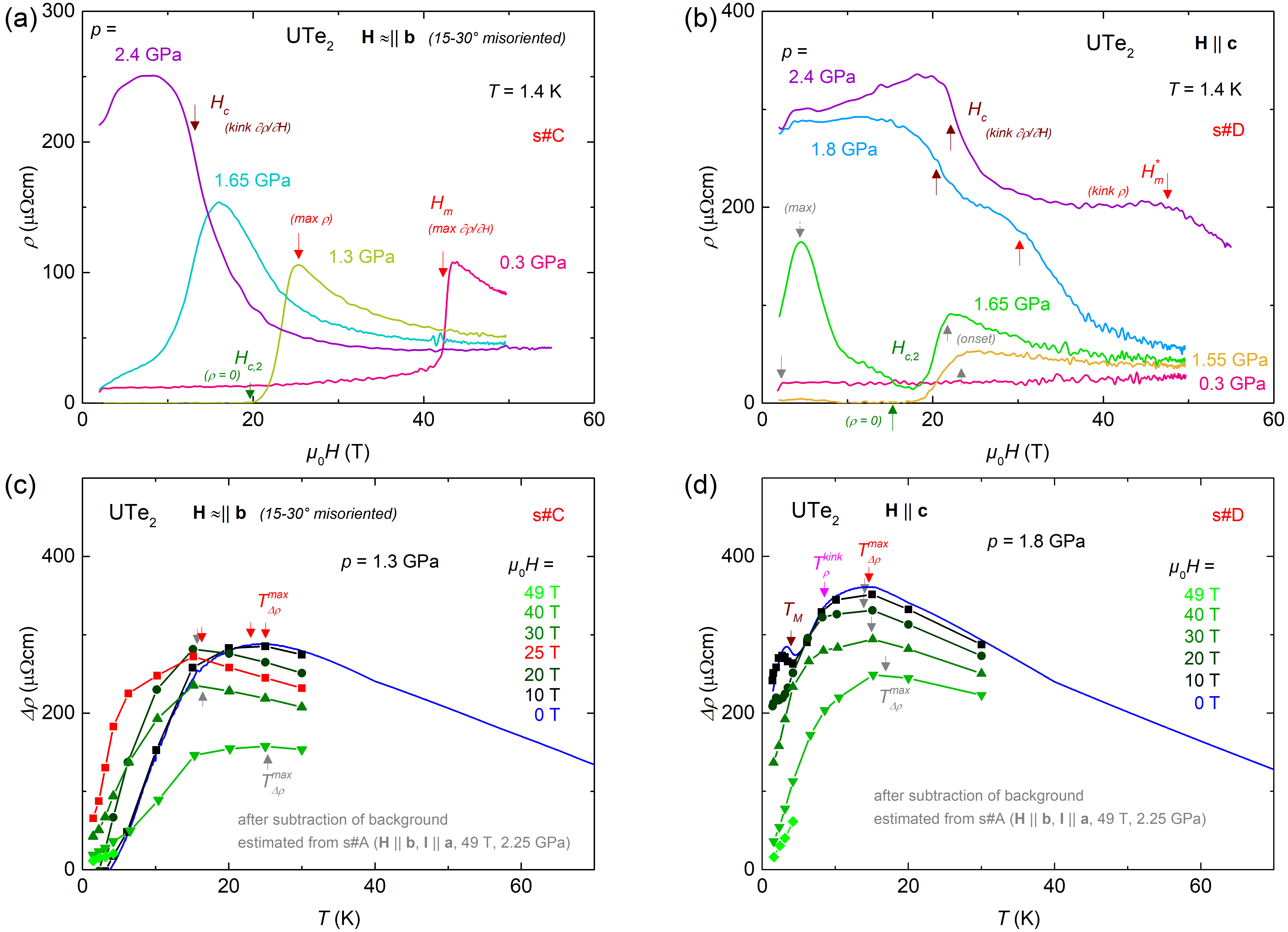}
\caption{Magnetoresistivity $\rho(H)$ curves at different pressures and at the lowest temperature (1.4~K) for the configurations (a) $\mathbf{H}\approx\parallel\mathbf{b}$ and (b) $\mathbf{H}\parallel\mathbf{c}$. Reconstructed temperature dependence of the resistivity $\Delta\rho$ determined after subtraction of the background under magnetic field (c) $\mathbf{H}\approx\parallel\mathbf{b}$ at 1.3~GPa and (d) $\mathbf{H}\parallel\mathbf{c}$ at 1.8~GPa (2b).}
\label{Fig3}
\end{figure}

In Fig. \ref{Fig3}(a-b) we show the magnetoresistivity curves at the lowest temperature (1.4~K) for different pressures and the two magnetic-field orientations. Most of the main results of this study are already apparent here. For the sample set up with $\mathbf{H}\parallel\mathbf{b}$, at low pressure the first-order metamagnetic transition appears as a huge and sharp increase of the resistivity, similar to what is seen at ambient pressure. However, here at 0.3~GPa this occurs at $\mu_0H_m\simeq43$~T, a field significantly higher than at ambient pressure (about 35~T) whereas it has previously been shown that $H_m$ decreases with pressure \cite{Knebel2020,Li2021}. The most likely explanation for this discrepancy is that the sample was somewhat misaligned in respect to the field, probably having moved inside the pressure cell on pressurization. Indeed, it has been shown that $H_m$ increases when the field is rotated away from the $b$-axis in both the $\mathbf{b}-\mathbf{c}$ and $\mathbf{b}-\mathbf{a}$ planes \cite{Ran2019b}. The value of $H_m$ found here would imply quite a large misalignment, between 15 and 30~$^\circ$. However, as the effect of a magnetic field up to 35~T applied along the $b$-axis has already been well studied \cite{Knebel2020}, we will see that this tilted configuration allows us to capture the essential physics for $\mathbf{H}\parallel\mathbf{b}$, as well as revealing interesting results for a field applied with some misalignment from the $b$-axis. In the following we will refer to this field configuration as $\mathbf{H}\approx\parallel\mathbf{b}$. On increasing pressure, the metamagnetic transition remains clear up to $p_c$ where $H_m$ decreases to about 12~T. At higher pressure the aspect of the curve changes and shows several features that we will explicit further on. For the sample with $\mathbf{H}\parallel\mathbf{c}$ the resistivity curve is basically featureless at 0.3~GPa, similar to the ambient pressure results \cite{Knafo2019}. In the high pressure curves (1.8 and 2.4~GPa), the zero field resistivity at 1.4~K has increased considerably, and shows a decrease with field in two steps with the corresponding field values marked here as $H_c$ and $H_m^*$. Concerning the superconductivity, the lowest temperature reached in this study (1.4~K) is only slightly below the ambient pressure superconducting critical temperature, so at low pressures we see almost no trace of superconductivity. However, $T_{sc}$ increases significantly with pressure, and our measurements give a good indication of the superconducting phase diagram. Two important effects are visible here. First for $\mathbf{H}\approx\parallel\mathbf{b}$, we see that at 1.3~GPa, superconductivity extends up to $\mu_0H_m\simeq20$~T. At ambient pressure, it has been shown that for a field perfectly aligned along the $b$-axis, superconductivity is reinforced with field and exists up to $H_m$, but that this effect disappears with a misalignment of just a few degrees, and $H_{c2}$ is considerably reduced. With the misalignment necessary to explain the large value of $H_m$ in our case we would certainly not expect superconductivity to extend up to $H_m$ at ambient pressure, so this implies that pressure strongly changes the phase diagram when the field is rotated away from the $b$-axis. The second remarkable effect is seen for $\mathbf{H}\parallel\mathbf{c}$, where at 1.55~GPa the resistivity is not zero at low fields, but a re-entrant superconducting state with zero resistivity is found at high field, between approximately 8 and 18~T, similar to the previous reports \cite{Ran2020,Aoki2021}.

\onecolumngrid

\begin{figure}[h]
\includegraphics[width=1\columnwidth]{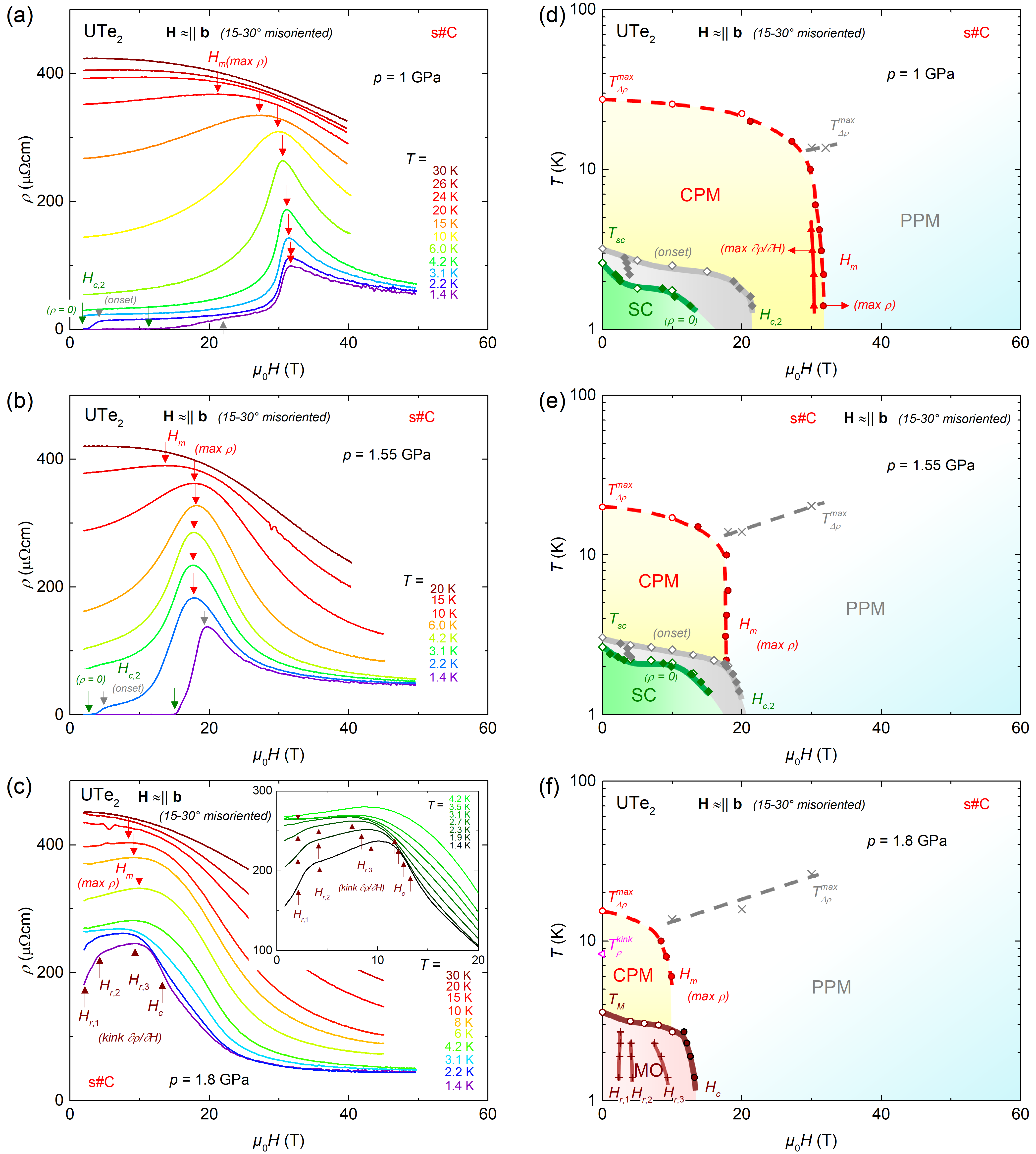}
\caption{Left-hand graphs: magnetoresistivity curves for the configuration $\mathbf{H}\approx\parallel\mathbf{b}$  at different temperatures for the pressures (a) $p=1$~GPa, (b) $p=1.55$~GPa, and (c) $p=1.8$~GPa. Right-hand graphs: obtained magnetic-field-temperature phase diagrams of the superconducting and magnetically ordered phases, and of the CPM regime delimited by $H_m$ and $T_{\Delta\rho}^{max}$, for the pressures (d) $p=1$~GPa, (e) $p=1.55$~GPa, and (f) $p=1.8$~GPa. CPM denotes the correlated paramagnetic regimes , PPM the polarized paramagnetic regime, SC the superconducting phase, and MO the magnetically-ordered phase.}
\label{Fig4}
\end{figure}

\begin{figure}[h]
\includegraphics[width=1\columnwidth]{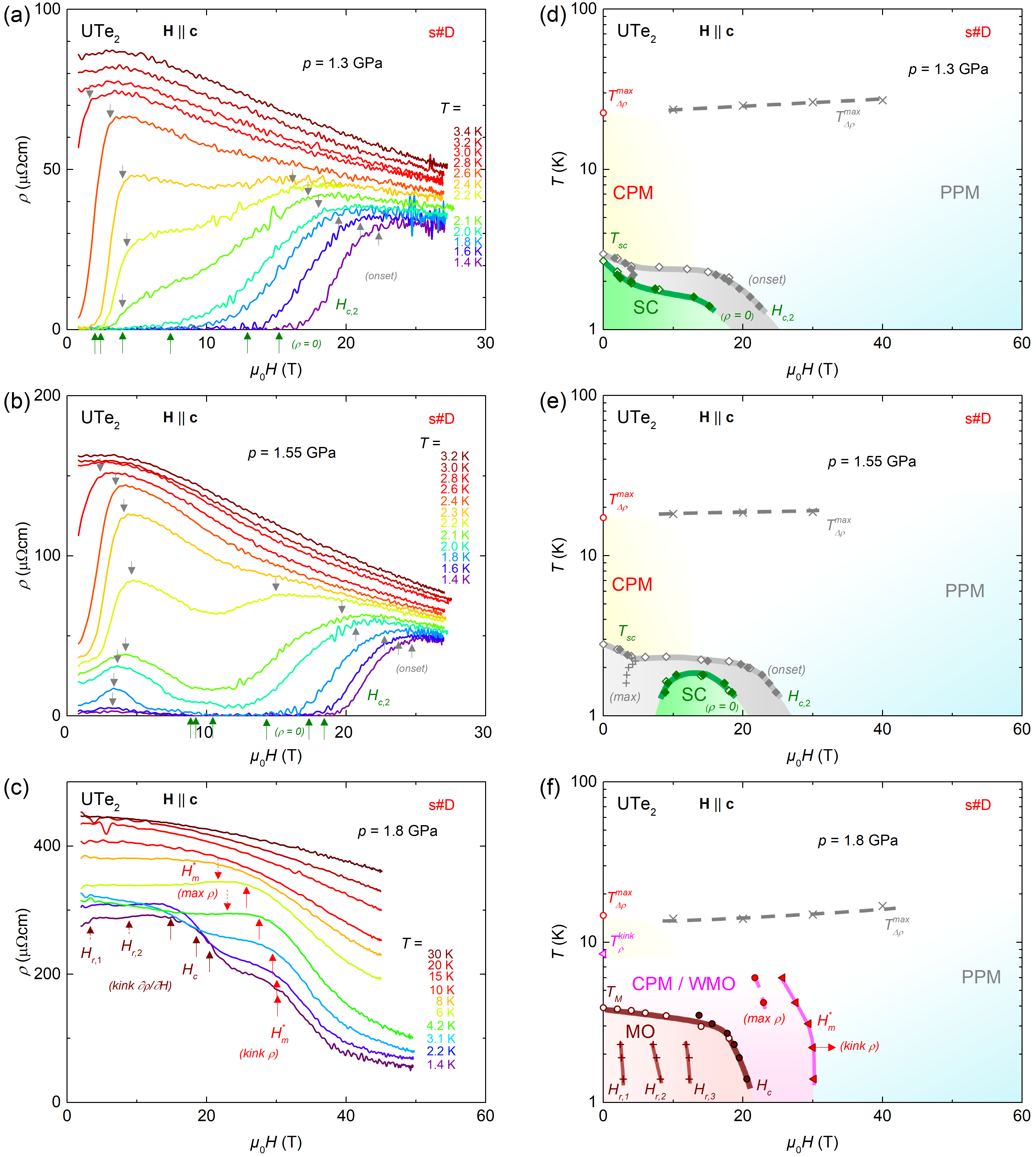}
\caption{Left-hand graphs: magnetoresistivity curves for the configuration $\mathbf{H}\parallel\mathbf{c}$  at different temperatures for the pressures (a) $p=1.3$~GPa, (b) $p=1.55$~GPa, and (c) $p=1.8$~GPa. Right-hand graphs: obtained magnetic-field-temperature phase diagrams of the superconducting and magnetically ordered phases, and of the CPM regime delimited by $H_m$ and $T_{\Delta\rho}^{max}$, for the pressures (d) $p=1.3$~GPa, (e) $p=1.55$~GPa, and (f) $p=1.8$~GPa. CPM and CPM/WMO denote the correlated paramagnetic regimes stabilized at pressures below and above $p_c$, PPM the polarized paramagnetic regime, SC the superconducting phase, and MO the magnetically-ordered phase.}
\label{Fig5}
\end{figure}

\begin{figure}[h]
\includegraphics[width=1\columnwidth]{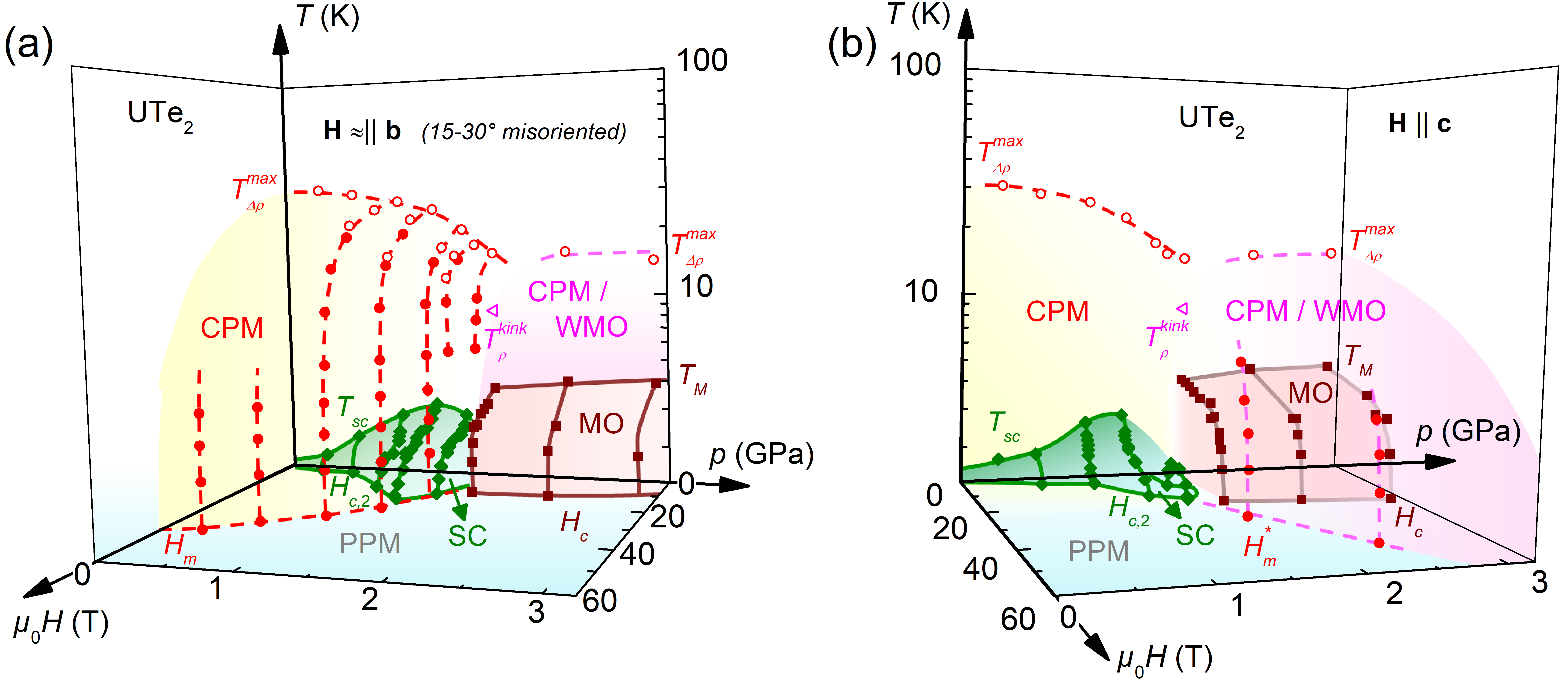}
\caption{3D magnetic-field-pressure-temperature phase diagrams for the configurations (a) $\mathbf{H}\approx\parallel\mathbf{b}$ and (b) $\mathbf{H}\parallel\mathbf{c}$. CPM and CPM/WMO denote the correlated paramagnetic regimes stabilized at pressures below and above $p_c$, PPM the polarized paramagnetic regime, SC the superconducting phase, and MO the magnetically-ordered phase.}
\label{Fig6}
\end{figure}

\begin{figure}[h]
\includegraphics[width=1\columnwidth]{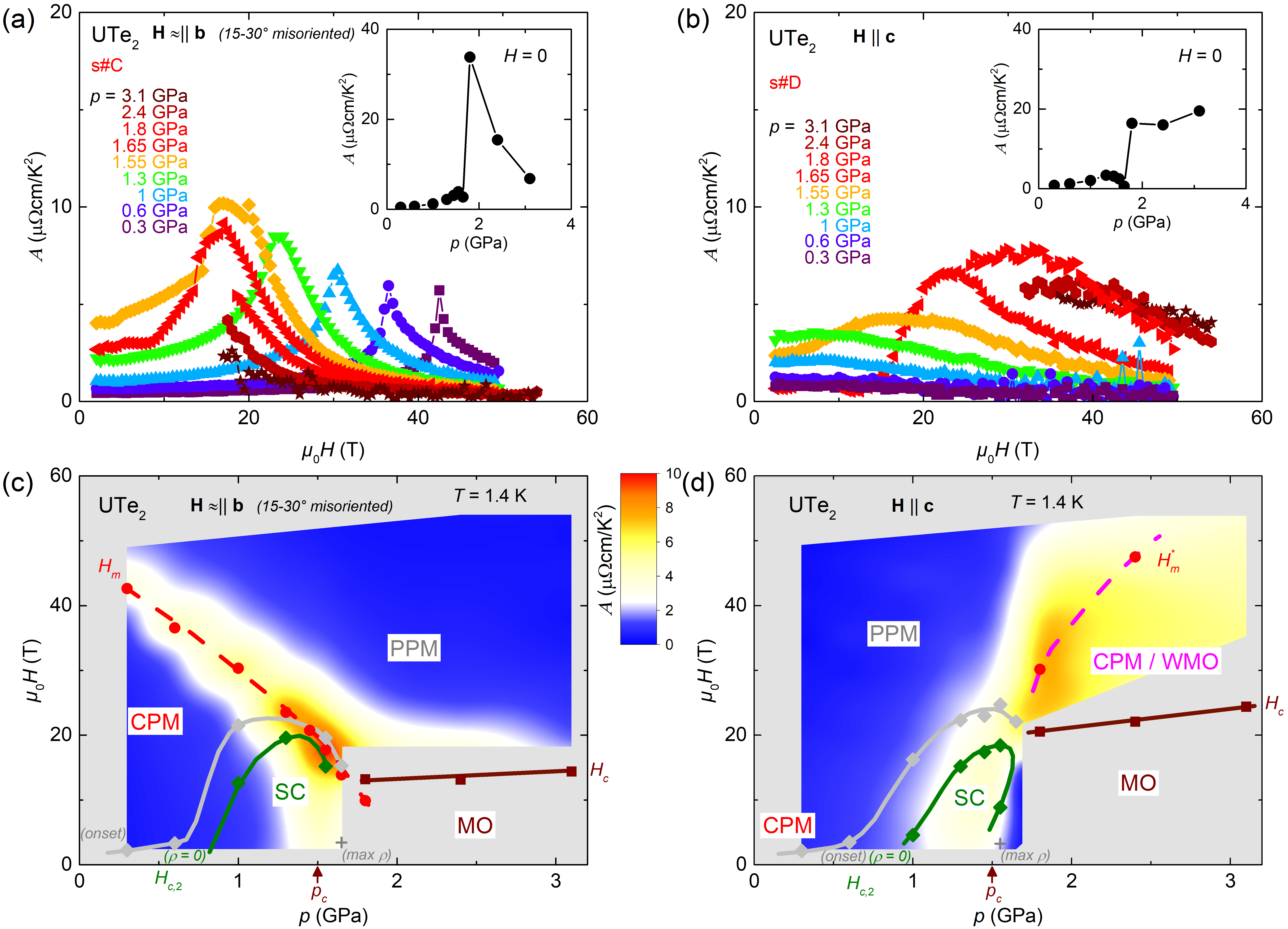}
\caption{Top graphs: Field dependence of the quadratic temperature dependence coefficient $A$ at different pressures for the configurations (a) $\mathbf{H}\approx\parallel\mathbf{b}$ and (b) $\mathbf{H}\parallel\mathbf{c}$. Lower graphs (c-d): colour plot of $A$ with superimposed pressure-magnetic-field phase diagrams showing how, for both configurations, the enhancement of superconductivity coincides with an enhancement of $A$. The coefficient $A$ was not determined within the grey regions.}
\label{Fig7}
\end{figure}

\twocolumngrid

A fuller understanding of the above effects and the complex phase diagram can be obtained by looking in detail at the temperature dependence of the magnetoresistivity. First, we examine the case for $\mathbf{H}\approx\parallel\mathbf{b}$. Fig. \ref{Fig4} (left-hand graphs) shows the resistivity curves for different temperatures and three pressures. For $p=1$ and 1.55~GPa, i.e. $p<p_c$, the curves are qualitatively similar to the ambient pressure results \cite{Knafo2019}. At low temperature, the first-order metamagnetic transition to the polarized paramagnetic (PPM) regime appears as a sharp and large increase of the resistivity. As temperature is increased this anomaly transforms into a broad maximum indicating a crossover delimiting the CPM regime. $H_m$ decreases with pressure as mentioned previously, and apart from the larger value of $H_m$ due to the misalignment of the field our results are similar to the previous study under pressure for $\mathbf{H}\parallel\mathbf{b}$ \cite{Knebel2020}. The phase diagrams drawn from these anomalies are shown in the right-hand panels of Fig. \ref{Fig4}. The field where the maximum of the resistivity occurs decreases as pressure is increased, similarly to the ambient pressure results \cite{Knafo2019}, and connects with the temperature $T_{\chi(\mathbf{H}\parallel\mathbf{a})}^{max}$ that also decreases with pressure \cite{Li2021}. We can see that the temperature $T_{\Delta\rho}^{max}$ at the maximum of the zero-field resistivity after subtraction of the background described above, which corresponds approximately to $T_\chi^{max}$ at zero field, also corresponds to $H_m$ in the range where both features can be seen. This implies that $T_{\Delta\rho}^{max}$ is indeed a good criterion to determine the boundary of the CPM regime. The superconducting phase diagrams are also plotted for $p=1$ and 1.55~GPa. For both pressures $H_{c2}$ shows an S-shape and, as already mentioned, at 1.55~GPa superconductivity extends up to $H_m$ despite the probable misalignment of field. These results will be discussed in more detail further on. The lower panels of Fig. \ref{Fig4} show the results for $\mathbf{H}\approx\parallel\mathbf{b}$ at 1.8~GPa. The high temperature magnetoresistivity curves retain the characteristic broad maximum seen at pressures below $p_c$, indicating a cross-over into a polarized state. This maximum disappears at temperatures above 10-15~K, corresponding to the value of $T_{\Delta\rho}^{max}$, and becomes a broad decrease of the resistivity with field. However, the magnetoresistivity curves at low temperature are quite different. Now $p>p_c$ and the ground state is almost certainly some kind of long range magnetic order below $T_M\simeq3$~K. For $T<T_M$ a pronounced kink can be seen at $\mu_0H_c\simeq13$~T corresponding to the transition from the long range magnetic order to the polarized paramagnetic state. Several other features are visible in the magnetoresistivity at the fields $\mu_0H_{r,1}=2.3$~T, $\mu_0H_{r,2}=4.3$~T, and  $\mu_0H_{r,3}=9.3$~T~$<\mu_0H_c$ at $T=1.4$~K. These transitions are presumably related to magnetic moment reorientations within the magnetically ordered phase (see Supplementary Materials \cite{SM}).

We now look at the case $\mathbf{H}\parallel\mathbf{c}$ (Fig. \ref{Fig5}). For this configuration no metamagnetic transition has been observed at ambient pressure at least up to 70~T \cite{Ran2019}. For the pressures 1.3 and 1.55~GPa, i.e., $p<p_c$, at high temperatures no particular feature is apparent in the magnetoresistivity. The most interesting result here concerns the superconductivity. At ambient pressure the $H_{c2}(T)$ curve for $\mathbf{H}\parallel\mathbf{c}$ shows no indication of enhancement of superconductivity with field. However several studies have already shown that under pressure the situation changes, with the slope of $H_{c2}$ becoming extremely steep \cite{Braithwaite2019,Knebel2020}, possibly indicating the appearance of a field-enhancement of superconductivity for $\mathbf{H}\parallel\mathbf{c}$, and re-entrant superconductivity appearing close to the critical pressure. These effects are confirmed here:  the upper and middle panels of Fig. \ref{Fig5} show that $\mu_0H_{c2}$ reaches about 20~T at 1.3~GPa and that re-entrant superconductivity develops at 1.55~GPa, where $\mu_0H_{c2}$ exceeds 20~T, respectively. The lower panels of Fig. \ref{Fig5} show that at 1.8~GPa, i.e., for $p>p_c$, the critical field $H_c$ where the low-temperature long-range magnetic order is destroyed can be seen as a well-defined kink in the curves for temperatures below $T_M\simeq3$~K. At the lowest temperature the critical field reaches $\mu_0H_c\simeq20$~T and it shifts slightly to lower field as the temperature increases. Similarly to the $\mathbf{H}\approx\parallel\mathbf{b}$ configuration, three anomalies in the electrical resistivity can be defined at the critical fields $\mu_0H_{r,1}=2.8$~T, $\mu_0H_{r,2}=8.1$~T, and  $\mu_0H_{r,3}=12.3$~T~$<\mu_0H_c$ at $T=1.4$~K, within the magnetically ordered state (see Supplementary Materials \cite{SM}). As for the configuration $\mathbf{H}\approx\parallel\mathbf{b}$, all these features disappear when the temperature is raised above $T_M$, confirming their link to the low-temperature magnetic order. However, for $p>p_c$, in contrast to the $\mathbf{H}\approx\parallel\mathbf{b}$ configuration a pronounced kink remains at higher field suggesting a transition or a well-defined cross-over into the polarized state. We denote this field $\mu_0H_m^*\simeq30$~T and speculate on its pseudo-metamagnetic nature, in analogy with the metamagnetic field where the polarized state occurs at low pressure for $\mathbf{H}\approx\parallel\mathbf{b}$, although here it does not show a sharp first-order transition. The nature of the regime between $H_c$ and $H_m^*$ is not clear. Anomalies have previously been seen in the resistivity, the specific heat and the magnetization \cite{Thomas2020,Li2021} at a temperature higher than $T_M$ for pressures close above $p_c$. Indeed, in the present study a clear kink can be seen in the zero-field resistivity at 1.8~GPa (Fig. \ref{Fig3}). It has been suggested that this phase could correspond to static \cite{Thomas2020} or short-range \cite{Li2021} weak magnetic order (WMO). The field $H_m^*$ probably corresponds to a transition or crossover between this phase and the polarized paramagnetic regime. Interestingly, a maximum in the magnetic susceptibility was observed at a temperature $T_\chi^{max}=11$~K for a pressure $p=1.8$~GPa~$>p_c$ and a magnetic field $\mathbf{H}\parallel\mathbf{c}$ \cite{Li2021}. Similarly to the low-pressure CPM regime delimited in a magnetic field $\mathbf{H}\parallel\mathbf{b}$ by $T_{\chi(\mathbf{H}\parallel\mathbf{b})}^{max}$ and $H_m$ \cite{Knafo2019,Miyake2019} (see also other heavy-fermion systems \cite{Aoki2013,Knafo2021c}), the WMO regime may also correspond to a second CPM regime delimited in a magnetic field $\mathbf{H}\parallel\mathbf{c}$ by $T_{\chi(\mathbf{H}\parallel\mathbf{c})}^{max}$ and $H_m^*$. In the following, we will label this regime as CPM/WMO.

\subsection{Magnetic quantum criticality and superconductivity}

The full 3D phase diagrams obtained for both configurations of magnetic field $\mathbf{H}\parallel\approx\mathbf{b}$ and $\mathbf{H}\parallel\mathbf{c}$ are represented in Fig. \ref{Fig6}. We see that the phase diagram for the magnetic order is quite similar for both configurations, with the ordering temperature being suppressed with field, but with a well-defined transition even at high field. A similar behavior has also been seen for $\mathbf{H}$ applied along the easy magnetic axis $\mathbf{a}$, where the critical field is even smaller \cite{Aoki2020}. The succession of field-induced transitions at fields $H_{r,i}< H_c$, with $i=1-3$, (presented in Fig. \ref{Fig4} and  \ref{Fig5}, but not in Fig. \ref{Fig6} for clarity) indicates the stabilization of different magnetic structures. Such behaviour would not be expected for a ferromagnetically-ordered phase, at least for a field applied along the easy axis, strongly suggesting that the magnetic order is of an antiferromagnetic (or spin-density-wave) type, as already inferred from previous results \cite{Braithwaite2019,Knebel2020,Thomas2020,Aoki2020}. A swich of polarization processes occurs at $p_c$ and leads to quite different 3D phase diagrams for the two configurations $\mathbf{H}\parallel\approx\mathbf{b}$ and $\mathbf{H}\parallel\mathbf{c}$:
\begin{itemize}
  \item At pressures below $p_c$, for $\mathbf{H}\approx\parallel\mathbf{b}$ the CPM regime appears as a well-defined 3D bubble, delimited by the first-order transition at $H_m$ at low temperature and the temperature $T_{\Delta\rho}^{max}$ (or $T_\chi^{max}$) at low field. However, for $\mathbf{H}\parallel\mathbf{c}$, while at zero field the cross-over between the low-temperature CPM and high-temperature PM regimes is obviously the same, there is no signature of transition to the PPM regime with field, and it is likely that a change develops smoothly as a continuous rotation of the moments. This is consistent with the absence of a maximum in the magnetic susceptibility for $\mathbf{H}\parallel\mathbf{c}$, which is almost Curie-Weiss-like down to the lowest temperatures \cite{Ikeda2006,Knafo2021}, and with the fact that $T_{\Delta\rho}^{max}$ increases with applied field for $\mathbf{H}\parallel\mathbf{c}$.
  \item For pressures above $p_c$ the situation is quite different. Now a 3D bubble formed by the CPM/WMO regime occurs for $\mathbf{H}\parallel\mathbf{c}$, with a quite well-defined transition into the PPM regime, whereas for $\mathbf{H}\approx\parallel\mathbf{b}$ this probably occurs as a broad cross-over. This is also consistent with the appearance of a maximum of magnetic susceptibility for $\mathbf{H}\parallel\mathbf{c}$ and the disappearance of such maximum for $\mathbf{H}\parallel\mathbf{b}$ under pressures beyond the critical pressure \cite{Li2021}.
\end{itemize}
This reshuffling of the magnetic properties at the critical pressure $p_c$ is related to a switch of the anisotropy of the magnetic susceptibility, with the hard magnetic axis $\mathbf{b}$ at ambient pressure becoming the easy magnetic axis at high-pressure \cite{Li2021}.

In heavy-fermion materials, magnetic quantum criticality, generally accompanied by a Lifshitz Fermi-surface instability, is reflected in the enhancement of the effective mass $m^*$ as the field is increased towards $H_m$ seen by a direct measurement of the specific heat \cite{Imajo2019}, as well as from magnetization \cite{Miyake2019} and resistivity \cite{Knafo2019} measurements. Magnetic fluctuations are often considered as the origin of the large effective mass $m^*$ observed in these materials \cite{Knafo2021c}. The quadratic temperature coefficient $A$, obtained by a fit of the resistivity to a Fermi-liquid behavior $\rho=\rho_0+AT^2$, varies as $m^{*2}$ within first approximation and shows a pronounced maximum at $H_m$ for $\mathbf{H}\parallel\mathbf{b}$, as well for a field tilted by 30~$^\circ$ in the $\mathbf{b}-\mathbf{c}$ plane \cite{Knafo2021}. In Fig. \ref{Fig7} we show the field dependence of the $A$ coefficient for different pressures, extracted from the reconstructed temperature dependences of the resistivity for both configurations. For the configuration $\mathbf{H}\approx\parallel\mathbf{b}$, at the lowest pressures the metamagnetic transition appears as a very sharp peak at $H_m$, starting at about 43~T at 0.3~GPa. As pressure is increased the peak position moves to lower fields, the value of $A$ increases, and above 1.3~GPa the peak starts to broaden noticeably. For the highest pressures ($p>p_c$) the low field ($H<H_c$) points are omitted as the onset of magnetic order occurring close to the lowest temperature measured here made the analysis meaningless in this case. For the configuration $\mathbf{H}\parallel\mathbf{c}$, at low pressure no feature is visible in $A(H)$. However, above 1~GPa a broad maximum becomes apparent that shifts to higher fields and becomes more pronounced as pressure is increased, remaining visible even for $p>p_c$.

The considerable changes of the magnetic properties with pressure have strong consequences on the superconductivity. In the 3D phase diagrams of Fig. \ref{Fig6}, an enhancement of superconductivity can be seen close to $p_c$ for both configurations of magnetic field $\mathbf{H}\approx\parallel\mathbf{b}$ and $\mathbf{H}\parallel\mathbf{c}$. Previous studies have shown that field re-entrant superconductivity develops at ambient pressure and low temperature for a sample perfectly aligned with $\mathbf{H}\parallel\mathbf{b}$ and that at a temperature of 1.4~K the sample should be superconducting at all fields up to $H_m$ once a small pressure is applied \cite{Knebel2019,Knebel2020}. Here, due to misalignment, no re-entrant superconductivity is seen for $\mathbf{H}\approx\parallel\mathbf{b}$ at low pressure and $H_{c2}$ is quite low at $T=1.4$~K. However, we find that superconductivity extends up to $H_m$ as pressure is increased. A similar result was also found recently for a sample oriented at an angle of 30~$^\circ$ from the $b$-axis in the $\mathbf{b}-\mathbf{c}$ plane \cite{Ran2021}. Concerning the configuration $\mathbf{H}\parallel\mathbf{c}$, we have evidenced the presence of field-induced superconductivity under pressures $p\lesssim p_c$. The lower panels (c-d) of Fig. \ref{Fig7} show the electronic $(p,H)$ phase diagrams at our base temperature, $T=1.4$~K, with the evolution of $A$ as a color plot. They emphasize the relationship between the enhancement of $A$ and the high-field stabilization of superconductivity. Field-reinforced or field-induced superconductivity is observed close to the critical pressure, where the collapse of the field scales $H_m$ and $H_m^*$ and enhanced $A$ coefficients are observed. For $\mathbf{H}\approx\parallel\mathbf{b}$ superconductivity survives up to $H_m$ under pressures near to $p_c$, where $A$ reaches its maximum value at $H_m$. For $\mathbf{H}\parallel\mathbf{c}$ and $p=1.55$~GPa, although no field-induced magnetic transition is observed and temperatures larger than $T_{sc}$, superconductivity may result from the proximity of critical magnetic fluctuations, as indicated by the enhancement of $A$ in a nearby region of the phase diagram. $A$ is maximum near $H_m^*$ for $p\gtrsim p_c$, and the field-induced superconducting phase which develops for $p\lesssim p_c$ appears as a prolongation of the $H_m^*$ line.

\section{Discussion}

Quantum criticality, either purely magnetic or accompanied by a Fermi-surface instability, is suspected to be a driving force for superconductivity in many heavy-fermion systems.  In UTe$_2$ this is evidenced by the enhancement of superconductivity on approaching the magnetic phase transition under pressure, and on approaching metamagnetic transitions with field. For the latter case, a quantitative analysis has shown that the re-entrant superconducting behaviour for $\mathbf{H}\parallel\mathbf{b}$ can be explained by a monotonic increase of the pairing strength related to the increase of the effective mass $m^*$ \cite{Knebel2019,Brison2021}. Furthermore, the rapid disappearance of the re-entrant behaviour as soon as the field is rotated away from the $b$-axis is a natural consequence, mainly due to the increase of $H_m$. Indeed, as the enhancement of the pairing strength occurs at higher fields, it is no longer sufficient to overcome the orbital and possibly paramagnetic pair-breaking effects at lower fields. A small anisotropy of $m^*$, that is maximum for $\mathbf{H}\parallel\mathbf{b}$ can further amplify this phenomenon. From this picture, it is easy to understand our result. As can be seen from the $A(H)$ curves, as pressure is increased the enhancement of $m^*$ and consequently of the pairing strength will simultaneously occur at lower fields and become stronger, allowing the field enhancement of superconductivity to be effective over a much wider angular range. We expect that, with a measurement made at lower temperatures, we would have found a superconducting state extending up to $H_m$ at even lower pressure. An open question is whether on approaching $p_c$ the field enhancement of superconductivity is still effective or not. In a previous study with $\mathbf{H}\parallel\mathbf{b}$, the characteristic S-shape of $H_{c2}$ was lost for pressures of 1~GPa and above \cite{Knebel2020}. The initial slope remained large but $H_{c2}$ showed a pronounced curvature that could be an indication of increasingly effective Pauli limitation with pressure. In the present study, for $\mathbf{H}\approx\parallel\mathbf{b}$ we see the S-shape of $H_{c2}$ at 1.55~GPa implying that the field reinforcement effect might still be active. This would be consistent with the pressure and field dependence of $A$, which still shows a pronounced maximum at $H_m$, and reaches higher values than at ambient pressure. What is clear from the study with configuration $\mathbf{H}\approx\parallel\mathbf{b}$ is that the metamagnetic transition at $H_m$ still acts as a very effective cut-off for superconductivity which does not survive in the polarized state. This is similar to the ambient pressure behavior for $\mathbf{H}\parallel\mathbf{b}$ but now concerns a wider angular range. It contrasts with the surprising re-entrant superconducting phase only seen in the PPM regime stabilized above $H_m$ in the configuration with a magnetic field $H$ tilted by 30~$^\circ$ from $\mathbf{b}$ toward $\mathbf{c}$ at ambient pressure \cite{Ran2019b,Knafo2021}. Recently a study performed in this configuration under pressure showed that superconductivity extends continuously below and above $H_m$ \cite{Ran2021}. It seems that here we are not in this configuration, implying that we probably have also a component of field along the $a$-axis, and so the extension of superconductivity up to $H_m$ may be found also for misalignement of the field in the $\mathbf{a}-\mathbf{b}$ plane under pressure.

For $\mathbf{H}\parallel\mathbf{c}$, the extremely steep slope found for $H_{c2}$ on approaching $p_c$, as well as the very high values ($25-30$~T) found for $H_{c2}$ here and in a previous study \cite{Aoki2021}, strongly suggest that pressure causes an enhancement of the superconducting pairing strength to come into play.  Fig. \ref{Fig7} shows clearly how this enhancement seems to be linked to the maximum of $A(H)$ that appears under pressure, and thus to the new field scale $H_m^*$, corresponding to the crossover to the polarized state from the CPM/WMO state. Approaching $p_c$ we find a re-entrant behaviour of superconductivity as has been reported previously \cite{Aoki2021}. While the field enhancement of the pairing strength is certainly favorable for this re-entrant behavior, the main ingredient is probably the competition between the magnetically ordered and superconducting phases. Indeed previous studies (in a magnetic field along $\mathbf{c}$ \cite{Aoki2021} or along an undetermined direction \cite{Ran2020}) showed that superconductivity only appears above the field necessary to suppress the magnetic order. This would suggest that in the present work the measurement at 1.55~GPa is actually at a pressure slightly above $p_c$.  For $\mathbf{H}\parallel\mathbf{c}$ and close to $p_c$, a question is whether superconductivity can develop in the correlated or polarized paramagnetic regimes, or in both.

Interestingly, the electrical resistivity $\rho$ measured here with a current $\mathbf{I}\parallel\mathbf{a}$ captures the physics driving the maxima in the magnetic susceptibility for different directions of magnetic field. Fig. \ref{Fig2} shows the similar values of $T_{\Delta\rho}^{max}$ and $T_{\chi(\mathbf{H}\parallel\mathbf{b})}^{max}$ for $p<p_c$, and of $T_{\Delta\rho}^{max}$ and $T_{\chi(\mathbf{H}\parallel\mathbf{a})}^{max}\simeq T_{\chi(\mathbf{H}\parallel\mathbf{c})}^{max}$ for $p\gtrsim p_c$ \cite{Li2021}. A transverse relationship between the electrical resistivity and the magnetic susceptibility anisotropies may be the consequence of an anisotropic Kondo hybridization between conduction and localized $f$ electrons in UTe$_2$. Rich information about the electronic interactions responsible for the magnetic fluctuations in UTe$_2$ in its normal non-superconducting phases may be accessed via a careful investigation of the anisotropy of the electrical resistivity, with different electrical-current directions (see [\onlinecite{Eo2021}]), under different magnetic-field directions possibly combined with pressure.

We have seen that the application of pressure and high magnetic field on UTe$_2$ leads to an extremely complex phase diagram with a complete reshuffling of the magnetic anisotropy and strong associated effects on superconductivity. This helps us understand the intricate relationship between superconductivity and magnetism in this system. The clear signature of the destruction of magnetic order with field, as well as the signature of several field induced transitions inside the magnetically ordered phase, show that this order may be of antiferromagnetic or SDW type. Different domains of stability and exclusion of superconductivity are found under pressure and magnetic fields. Critical magnetic fluctuations, possibly of ferromagnetic kind, associated with $H_m$ and $H_m^*$, may induce the large values of $A$ observed here, in relation with the enhancement of the superconducting pairing mechanism. However, superconductivity does not necessarily occur in all parts of the phase diagram where $A$ is enhanced. Quite simple pictures can explain some parts of the phase diagram, but understanding why superconductivity is destroyed below or beyond a metamagnetic field, depending on the field direction, and in the magnetically-ordered phase stabilized under pressure remains a theoretical challenge. Experimentally, a full knowledge of the angle dependence of superconductivity under pressure and very high magnetic field would help gain a full understanding of superconductivity in UTe$_2$. Our study is a step in this direction.

\section*{Acknowledgements}

We thank J. Flouquet and J.-P. Brison for useful discussions. We acknowledge financial support from the Cross-Disciplinary Program on Instrumentation and Detection of CEA, the French Alternative Energies and Atomic Energy Commission, KAKENHI (JP15H05882, JP15H05884, JP15K21732, JP16H04006, JP15H05745, JP19H00646,JP19K03736) and GIMRT (19H0416, 19H0414), and the French national research agency Programme Investissements d'Avenir under Program No. ANR-11-IDEX-0002-02, Reference No. ANR-10-LABX-0037-NEXT, and  the collaborative research project FRESCO.


%

\end{document}